\documentclass[aip,jcp,,preprint,letterpaper]{revtex4}
\usepackage[utf8]{inputenc}
\usepackage{comment}
\usepackage{multirow}
\usepackage{array}
\usepackage{bm}
\usepackage{amsmath}
\usepackage{commath}
\usepackage{longtable}
\usepackage{color}
\usepackage{ textcomp }
\usepackage{diagbox}
\usepackage[utf8]{inputenc}
\usepackage{bm}
\usepackage{graphicx}
\usepackage{tikz}
\usetikzlibrary{quantikz}
\usepackage{braket}
\usepackage{float}
\usepackage{bbold}
\usepackage{appendix}

\newcommand{\id}{ \mathbb{1}}

\begin{document}

\title{Variational Quantum Simulation of Chemical Dynamics with Quantum Computers}
\author{Chee-Kong Lee}
\email{cheekonglee@tencent.com}
\affiliation{Tencent America, Palo Alto, CA 94306, United States}
\author{Chang-Yu Hsieh}
\affiliation{Tencent, Shenzhen, Guangdong 518057, China}
\author{Shengyu Zhang}
\affiliation{Tencent, Shenzhen, Guangdong 518057, China}
\author{Liang Shi}
\email{lshi4@ucmerced.edu}
\affiliation{Chemistry and Biochemistry, University of California, Merced, California 95343, United States}

\begin{abstract}
Classical simulation of real-space quantum dynamics is challenging due to the exponential scaling of computational cost with system dimensions. 
Quantum computer offers the potential to simulate quantum dynamics with polynomial complexity;
however, existing quantum algorithms based on the split-operator techniques require large-scale fault-tolerant quantum computers that remain elusive in the near future.
Here we present variational simulations of real-space quantum dynamics suitable for implementation in Noisy Intermediate-Scale Quantum (NISQ) devices. The Hamiltonian is first encoded onto qubits using a discrete variable representation (DVR) and binary encoding scheme. We show that direct application of real-time variational quantum algorithm based on the McLachlan's principle is inefficient as the measurement cost grows exponentially with the qubit number for general potential energy and extremely small time-step size is required to achieve accurate results. 
Motivated by the insights that most chemical dynamics occur in the low energy subspace, we propose a subspace expansion method by projecting the total Hamiltonian, including the time-dependent driving field, onto the system low-energy eigenstate subspace using quantum computers, the exact quantum dynamics within the subspace can then be solved classically. We show that the measurement cost of the subspace approach grows polynomially with dimensionality for general potential energy. Our numerical examples demonstrate the capability of our approach, even under intense laser fields. 
Our work opens the possibility of simulating chemical dynamics with NISQ hardware.
\end{abstract}
\maketitle

\section{Introduction}
Theoretical calculations of quantum dynamics in atoms and molecules are critical for our understanding of chemical processes and designing the optimal control for chemical reactions, among many other applications~\cite{Miller2005, Gatti2017}. However, despite many decades of tremendous advances in method development, accurate and efficient simulation of quantum dynamics remains one of the most challenging scientific endeavors in physics and chemistry due largely to the curse of dimensionality.  State-of-the-art approaches, such as the multiconfigurational time-dependent Hartree (MCTDH) method, can treat quantum systems with tens of degrees of freedom, but additional  approximations are necessary~\cite{Meyer2009}. 
Quantum computers offer the possibility to simulate real-time dynamics beyond the reach of classical computers with polynomial complexity in computer memory and execution time. 

Indeed, one of the founding concepts of quantum computers arises from Feynman’s idea of simulating  quantum many-body dynamics using another quantum system~\cite{Feynman1982}.  
Subsequently, Lloyd~\cite{Lloyd1996} proved that quantum computers can simulate quantum systems using resources that scale only polynomially with system sizes, as compared with the exponential scaling on classical computers. Wiesner~\cite{Wiesner1996} and Zalka~\cite{Zalka1998} suggested the first proposals using quantum computers for simulating the quantum dynamics in real space representation. Subsequently,  Kassal \textit{et. al.}~\cite{Kassal2008} introduced a general procedure how to implement the time evolution propagator of a real-space Hamiltonian in a quantum circuit. Benenti and Strini~\cite{Benenti2008} and Somma~\cite{Somma2016} presented a detailed implementation of the potential energy terms for a one-dimensional harmonic potential. More recently, quantum circuits for spin-boson and Marcus models have also been devised~\cite{Macridin2018, Ollitrault2020}.

The mentioned quantum algorithms rely on the existence of fault-tolerant quantum computers to outperform their classical counterparts, and therefore are too resource-intensive for near-term quantum hardware. Given the limits of current quantum devices in terms of qubit numbers and hardware errors, there are intense interests in the exploration and development of variational quantum algorithms (VQAs) for quantum simulations~\cite{Bharti2021, Cerezo2021}. 
Within a VQA, the wavefunction is represented by a parametrized quantum state that can be prepared efficiently in a quantum circuit, and the variational parameters are updated iteratively by classical computers via an optimization loop. 
As a result of the integration between quantum and classical computers, the VQA algorithm might be implemented with quantum circuits of much shallower depth compared with other conventional fault-tolerant algorithms for Hamiltonian simulations.

However, it remains unclear if chemical dynamics in real space can be simulated using VQA efficiently, particularly how to represent the Hamiltonian suitable for VQA without incurring prohibitive quantum measurement cost arising from the large number of grid points that grows exponentially with dimensionality. In this work we demonstrate that such variational simulation task is possible by using a discrete variable representation (DVR) of the Hamiltonian. 
With the Hamiltonian in the DVR basis and a low-energy subspace expansion method, we show that the measurement cost only grows polynomially with system dimensionality.

A straightforward approach is to simply map the DVR Hamiltonian to qubits using the binary encoding scheme and propagate the dynamics with the real-time VQA based on the McLachlan’s principle~\cite{Li2017a}.
 We show that while direct application of the real-time VQA can produce near exact chemical dynamics in principle, the algorithm could be unstable and requires a very small time step in order to produce reliable results. This could pose a challenge for NISQ hardware in which quantum resources are limited and errors cannot be entirely mitigated.
 Additionally, we show that in general the real time VQA is not efficient in terms of measurement cost due to the need to compute the expectation values of an exponential number of terms in the potential energy operator, $V$, after mapping to qubits. 
 
 Given the intuition that many chemical dynamics only involve a small number of low-energy quantum states, we propose projecting the system Hamiltonian onto low-energy subspace using quantum computers and compute the dynamics within the subspace classically. This can be done by first computing the low energy eigenstates of the (time-independent) system Hamiltonian either by imaginary time VQA or the standard variational quantum eigensolver (VQE), and then projecting the total Hamiltonian (including time-dependent driving fields) into the subspace spanned by these eigenstates.
 Given that the dimension of the subspace is much smaller than the total Hilbert space, the dynamics within the subspace can be solved exactly with classical computers.
 Additionally, such approach is general in the sense that it is applicable for any form of the potential energy. In comparison, methods based on Suzuki-Trotter decomposition relies on efficient implementation of the propagator $\mbox{e}^{-iVt}$ where $V$ is the potential energy. However such quantum circuits are only known in several special cases, e.g. one dimensional harmonic potential. 
We numerically demonstrate our approach with two applications (1) the isomerization reaction dynamics of malonaldehydes molecule and (2) the electron dynamics and harmonic generation spectrum in a one-dimensional model helium atom.
 We show that our approach agrees well with exact calculations using just a small number of low energy eigenstates, even under intense laser fields.

The paper is structured as follows: we first review the split-operator method for quantum dynamics in Sec.~\ref{sec:split_operator}. 
In Sec.~\ref{sec:dvr}, we show that Hamiltonian in real space representation can be mapped to qubits using the DVR basis from Ref.~\cite{Colbert1992} and a binary encoding scheme. 
In Sec.~\ref{sec:vqa}A we review the real-time VQA based on the McLachlan’s principle and in Sec.~\ref{sec:vqa}B we propose simulating the chemical quantum dynamics using a low-energy subspace expansion method. 
We provide quantum resource estimations in Sec.~\ref{sec:resource}, and show that the real-time VQA is not efficient in terms of measurement cost for general potential energy, whereas measurement cost for the low-energy subspace method only grows polynomially with dimensionality.
In Sec.~\ref{sec:numerics}, we provide two numerical examples, a one-particle problem of isomerization dynamics in a double-well potential, and a two-particle problem of electron dynamics in a one-dimensional model helium atom. 
Finally in Sec.~\ref{sec:discussion} we provide additional discussions on our approach and conclude.

\section{Quantum dynamics with split-operator method}\label{sec:split_operator}
We first review how quantum dynamics in real space representation can be simulated by fault-tolerant quantum computers using the split operator method. 
For simplicity, we consider one dimensional systems whose Hamiltonian in Cartesian coordinate is given by 
\begin{eqnarray}
H = T + V(x).
\end{eqnarray}
where $T$ and $V$ are the kinetic and potential operators, respectively. A general framework was first provided by Ref.~\cite{Wiesner1996, Zalka1998, Kassal2008} by first discretizing the space into $L$ grid points and encode the wavefunction onto $\log_2(L)$ qubits (see Sec. \ref{sec:dvr} for details on the encoding method). The wavefunction at time $t$ can then be obtained by employing the split operator technique 
and quantum Fourier transform (QFT)
\begin{eqnarray}
    \ket{\Psi(t)} = \mbox{e}^{-i H t} \ket{\Psi(0)} &\approx& \Big( \mbox{e}^{-i T t/n} \mbox{e}^{-i V t/n}\Big)^n  \ket{\Psi(0)}, \nonumber\\
    &=& \Big( F^{-1}\mbox{e}^{-i \frac{p^2}{2m} t/n} F \mbox{e}^{-i V t/n}\Big)^n  \ket{\Psi(0)}, \label{eq:trotter}
\end{eqnarray}
where $\mbox{e}^{-i T t/n} = F^{-1} \mbox{e}^{-i \frac{p^2}{2m} t/n}  F$ and $F$ is the QFT operator. The QFT can efficiently transform the wavefunction between the real-space and momentum-space representations, such that 
the potential energy term is diagonal in the coordinate representation and the momentum term is diagonal in the momentum representation. 

The quantum simulation is then reduced
to the implementation of the QFT plus diagonal operators of the form 
\begin{eqnarray} \label{eq:diagonal_trans}
    \ket{x} \longrightarrow \mbox{e}^{if(x)}\ket{x}.
\end{eqnarray}
However to the best of our knowledge, quantum circuits that efficiently implement Eq.~(\ref{eq:diagonal_trans}) (i.e. polynomial quantum resources with the number of qubits) have only been devised for functions quadratic in $x$ (i.e. one dimensional harmonic potential)\cite{Benenti2008}, spin-boson model~\cite{Macridin2018}, and Marcus model (coupled harmonic oscillators)~\cite{Ollitrault2020}. It remains unclear how to implement a general potential energy term in a quantum circuit efficiently.

\section{Hamiltonian in DVR representation} \label{sec:dvr}
In order to make quantum dynamics simulation amenable for variational simulation, we first express the real-space Hamiltonian within the discrete variable representation (DVR) proposed by Colbert and Miller~\cite{Colbert1992}. 
In this DVR, the one-dimensional kinetic and the potential energy terms in the grid basis can be written as 
\begin{eqnarray}
    T_{i,i'} &=& \hbar^2 (-1)^{i-i'}/ (2m\Delta x^2) 
    \begin{dcases}
        \frac{\pi^2}{3},& \text{if } i = i'\\
        \frac{2}{|i -i'|^2},  & \text{if }  i \neq i' 
    \end{dcases}\\
    V_{i,i'} &=& \delta_{ii'} V(x_i),
\end{eqnarray}
where $\Delta x$ is the grid spacing of the uniformly spaced grid points. $V(x_i)$ is the potential energy evaluated at position $x_i$. 

The $d$-dimensional generalization of DVR in Cartesian coordinates $\{x_1,x_2, ..., x_d\}$ is straightforward since the momentum operators are not coupled:
\begin{eqnarray}
H = \sum_{\alpha=1}^d T_\alpha + V,
\end{eqnarray}	
where 
\begin{eqnarray}
T_\alpha &=& \id_1 \otimes \id_2 ... \otimes \id_{\alpha-1} \otimes T  \otimes \id_{\alpha+1} \otimes  ...  \otimes \id_{d-1} \otimes \id_d, \\
V &=& \sum_{i_1,\cdots,i_d=1}^L V(x_{i_1},\cdots,x_{i_d}) | i_1 \cdots i_d \rangle \langle i_1 \cdots i_d |,
\end{eqnarray}
where $L$ is the number of grid points per dimension and the index $i_k$ denotes the $i$-th grid point in dimension $k$.  
It is worth noting that the Hamiltonian in DVR is sparse for higher dimensional systems, since the potential energy operator is diagonal while the kinetic energy operator is simply a sum of one-dimensional momentum operators. This is critical because the number of quantum measurements required depends heavily on the number of non-zero off-diagonal elements in the Hamiltonian (see Sec.~\ref{sec:resource} for more discussions).

In many cases (including our numerical examples), the system is initially prepared in the ground state or one of the eigenstates, and subsequently  perturbed by time-dependent driving fields in the form of laser pulses. Thus for the remaining of the paper, we write the total Hamiltonian in the form:
\begin{eqnarray}
    H = H_0 + H_I(t),
\end{eqnarray}
where $H_0$ is the molecular Hamiltonian and $H_I(t)$ denotes the time-dependent perturbation (e.g. laser pulses).

To express the DVR Hamiltonian in Pauli matrices, we use a standard binary encoding scheme in which $L$ discretized position basis states can be encoded in the quantum states of  $N = \log_2(L)$ qubits. For example, the quantum state representing grid point $m$, $\ket{x_m}$, can be represented by
\begin{eqnarray} \label{eqn:state_mapping}
\ket{x_m} = \ket{\vec{k}} = \ket{k_1} \otimes \ket{k_{2}} \otimes. . . \ket{k_{N}},
\end{eqnarray}
where the subscript denotes the qubit number and $m = k_1 2^0 + k_2 2^1 ... +k_{N}2^{N-1}$, and $k_i$ can be $0$ or $1$. The operator, $\ket{x_m}\bra{x_n}$, can be mapped to the qubit representation
\begin{eqnarray} 
\label{eqn:operator_mapping}
\ket{x_m}\bra{x_n} = \ket{\vec{k}}\bra{\vec{k}'} =  \ket{k_1}\bra{k'_{1}} \otimes \ket{k_2}\bra{k'_{2}} \otimes ... \otimes\ket{k_{N}}\bra{k'_{N}}, 
\end{eqnarray}
where each binary projector can be expressed in terms of Pauli matrices as follows
\begin{eqnarray}
\ket{0}\bra{1} = \frac{1}{2}( \sigma^x +  i  \sigma^y )&;&
\ket{1}\bra{0} = \frac{1}{2}( \sigma^x -  i  \sigma^y ); \nonumber \\
\ket{0}\bra{0} = \frac{1}{2}( I +   \sigma^z )&;& 
\ket{1}\bra{1} = \frac{1}{2}( I -   \sigma^z ).
\end{eqnarray}
where $I$ is the identity matrix, and $\sigma_x$, $\sigma_y$, and $\sigma_z$ are Pauli matrices. 
With the binary encoding scheme, a one-dimensional DVR Hamiltonian with $L$ grid points can be mapped to a qubit Hamiltonian of $N = \log_2(L)$ interacting qubits. For a $d$-dimensional system with $L$ grid points per dimension, the number of qubits required would simply be $N = d\log_2(L)$. 

\section{Variational Quantum Algorithms} \label{sec:vqa}
\subsection{Method 1: Direction Implementation of Real-Time VQA}
We first consider the real-time VQA introduced in Ref.~\cite{Li2017a} to simulate the chemical dynamics and assess its performance.
The time-dependent quantum state, $\ket{\Psi(t)}$, is approximated by a parametrized quantum state, $\ket{\psi(\vec \theta(t))}$, i.e.
$\ket{\Psi(t)} =  \hat{T} \mbox{e}^{-i \int^t_0 {H}(t') dt' } \ket{\Psi_0} \approx \ket{\psi( \vec \theta(t))}$ where 
$\vec \theta(t) = [\theta_1(t), \theta_2(t), \theta_3(t), ...] $ denotes the variational parameters at time $t$ and $\hat{T}$ is the time-ordering operator. According to McLachlan’s principle, the equation of motion for the variational parameters is obtained by minimizing the quantity $\|\Big (i\frac{\partial}{\partial t} -  H(t) \Big) \ket{\psi(\theta}\|$ which results in
\begin{eqnarray} \label{eq:EOM_theta}
\vec \theta(t + \delta t) = \vec \theta(t) +  \dot{\vec \theta} (t) \delta t; \,\,\, \dot{\vec \theta} (t) =  \mbox{Re} [ M^{-1}] \mbox{Im} [{f}], 
\end{eqnarray}
where the matrix elements of $ M$ and $ f$ are
\begin{eqnarray} \label{eq:EOM}
 M_{kl} =   \left\langle \frac{\partial \psi(\vec{\theta})}{\partial \theta_k} \right\vert \left. \frac{\partial \psi(\vec{\theta})}{\partial \theta_l}
\right\rangle; 
 f_{k} =   \left\langle  \psi(\vec{\theta}) \right\vert  H \left\vert \frac{\partial \psi(\vec{\theta})} {\partial \theta_k }
\right\rangle.  
\end{eqnarray}
The above matrix elements can be computed using the standard Hadamard test (see Appendix).

For analysis we consider a fairly general wavefunction ansatz of the form
\begin{eqnarray} \label{eqn:wfn_ansatz}
\ket{\psi(\vec{\theta)}} = {U}(\vec \theta)\ket{\psi_0}  = \prod_{k=1}^{N_{\theta}} {U}_k(\theta_k) \ket{\psi_0}  =  \prod_{k=1}^{N_\theta} \mbox{e}^{i \theta_k {R}_k}  \ket{\psi_0}. \label{eq:ansatz}
\end{eqnarray}
where $\ket{\psi_0}$ is the initial state of the wavefunction, ${R}_k$ is some Pauli string and $N_{\theta}$ denotes the number of variational parameters. 
The circuit depth for preparing such ansatz is therefore proportional to the number of variational parameters.  
It is worth noting that the methodology developed in our work is general and does not depend on the specific form of wavefunction ansatz.

In our numerical calculations in Sec.~\ref{sec:numerics}, we use a Hamiltonian variational ansatz (HVA) of two layers\cite{Wecker2015} and only include one- and two- qubits rotations. The qubit Hamiltonian after mapping could contain many-body terms encompassing all $N$ qubits, but such many-body terms are difficult to implement in quantum hardware and therefore are not included in our HVA. 
The HVA has been shown to exhibit favorable structural properties such as mild or entirely absent barren plateaus~\cite{Wiersema2020}, and its expressive power can be systematically improved by adding more layers. 

\subsection{Method 2: Low-Energy Subspace Expansion}
One disadvantage of the real-time VQA above is that the algorithm could be unstable and errors accumulate quickly, thus a very small time step is required for accurate long-time simulation (see Fig.\ref{fig:1D_VQA}(c) for example). Given that many of the molecular quantum dynamics only involve a small number of low-energy states, we propose expressing the total system Hamiltonian in the low-energy eigenstate subspace, and solving the quantum dynamics exactly within the reduced subspace with classical computer. Similar subspace methods are popular in computing the low-energy excited states of static Hamiltonian~\cite{McClean2017, Parrish2019, Nakanishi2019}. 

First we utilise quantum computers to compute the $N_s$ lowest eigenstates of the time-independent Hamiltonian, $H_0$. The $k$-th eigenstate can be found by computing the ground state of the modified Hamiltonian~\cite{higgott2019variational}
\begin{eqnarray} \label{eq:excited_hamiltonian}
    H_k = H_0 + \sum_{i=0}^{k-1} \beta_i \ket{\psi_i (\vec \theta) }\bra{\psi_i (\vec \theta)}, 
\end{eqnarray}
where $\ket{\psi_i (\vec \theta)}$ is the approximate $i$-th eigenstate found from previous calculations. Note that the penalty parameters should be large enough such that $\beta_i > E_k - E_i$ where $E_i$ is the energy of the $i$-th eigenstate. A detailed discussion on how to choose $\beta_i$ can be found in Ref.~\cite{higgott2019variational}. In our numerical examples, we use an imaginary-time VQA to compute the ground state of $H_k$~\cite{McArdle2019}. Similar to the real-time algorithm in Eq.~(\ref{eq:EOM_theta}) and (\ref{eq:EOM}), the update rule for the imaginary time algorithm is 
\begin{eqnarray} \label{eq:EOM_theta_imag}
\vec \theta(\tau + \delta \tau) = \vec \theta(\tau) +  \dot{\vec \theta} (\tau) \delta \tau; \,\,\, \dot{\vec \theta} (\tau) =  \mbox{Re} [ M^{-1}] \mbox{Re} [{f}], 
\end{eqnarray}
where $\delta \tau$ denotes the imaginary time step. 

After computing the required eigenstates of $H_0$, we express the total Hamiltonian, $H(t)$, in the subspace of these eigenstates, the matrix elements within the subspace become 
\begin{eqnarray}
    \tilde H_{ij} =  \bra{\psi_i (\vec \theta)} H \ket{\psi_j (\vec \theta)}. \label{eq:projection}
\end{eqnarray}
The above expression is also executed with quantum computers since it involves manipulating wavefunction of size $2^N$. The dimension of subspace Hamiltonian $\tilde H$ corresponds to the number of eigenstates we include, a number that should be much smaller than the dimension of the full Hilbert space. Finally we can solve the dynamics $\ket{\tilde \psi(t)} = \hat {T} \mbox{e}^{-i \int^t_0 \tilde H(s) ds} \ket{\tilde{\psi}(0)}$ via numerical integration. 

To evaluate the overlap $|\langle \psi_i(\vec \theta)| \psi_j (\vec \theta) \rangle|^2$ when computing the ground state of $H_k$, one can write the overlap as  $|\bra{\psi_0} U^\dagger_i(\vec \theta) U_j( \vec \theta) \ket{\psi_0}|^2$. In other words, the overlap can be computed by preparing the state $U^\dagger_i(\vec \theta) U_j( \vec \theta) \ket{ \psi_0}$ and perform measurement in $\ket{\psi_0}$ basis (in our simulations we use $\ket{\psi_0} = \ket{+}$). Thus the implementation of this method requires twice the circuit depth as compared to the state preparation of $\ket{\psi(\vec \theta)}$. 
An alternative approach involves the destructive SWAP gates and requires $2N$ qubits and $O(1)$ additional circuit depth~\cite{higgott2019variational}.

Next we discuss how to evaluate $\bra{\psi_i (\vec \theta)} H \ket{\psi_j (\vec \theta)}$ efficiently. 
The matrix elements for the unperturbed Hamiltonian is straightforward since $\bra{\psi_i (\vec \theta)} H_0 \ket{\psi_j (\vec \theta)} = \delta_{ij} E_i$ where $E_i$ is the energy of the $i$-th eigenstate.  We limit the time-dependent term to the form $H_I(t) = -\mu \varepsilon(t)$ (i.e. dipole approximation ) where $\mu$ is the dipole operator, and $\varepsilon(t)$ is the external electric field. Take a particle with charge $q$ in one dimension as an example, $\mu = q x$. Given the fact that $\mbox{e}^{i \epsilon x }$ can be implemented in linear time~\cite{Ollitrault2020}, $\bra{\psi_i (\vec \theta)} x \ket{\psi_j (\vec \theta)}$ can be computed using the following decomposition 
\begin{eqnarray}
    \bra{\psi_i (\vec \theta)} x \ket{\psi_j (\vec \theta)} = \lim_{\epsilon \to  0}\frac{i}{2\epsilon} \bra{\psi_i(\vec \theta)} \mbox{e}^{-i\epsilon x} -  \mbox{e}^{i\epsilon x} \ket{ \psi_j (\vec \theta)}.
\end{eqnarray}
Generalization to higher dimensions is straightforward.

\section{Resource estimations} \label{sec:resource}
Here we provide some estimates on the quantum resource requirements for the variational quantum simulation of the real-space quantum dynamics. 
Since the circuit depth for preparing the variational 
wavefunction in Eq.~(\ref{eq:ansatz}) is simply proportional to the number of variational parameters, $N_\theta$, our analysis mainly focuses on the measurement cost. 
Note that for the HVA used in our numerical examples, we only include terms up to two-qubit rotations, thus the number of variational parameters scales quadratically with the number of qubits, i.e. $N_\theta \propto O(N^2)$.

With the real and imaginary time VQAs, an independent circuit is needed for each matrix element $M_{ij} = \langle \frac{\partial \psi}{\partial \theta_i}| \frac{\partial \psi}{\partial \theta_j} \rangle $, so the estimation of the entire $M$ matrix would require $O(N_\theta^2)$ circuits. 
Estimating the force vector, $f_i = \langle \psi | H |  \frac{\partial \psi}{\partial \theta_i} \rangle $ is more involved.  We first consider the cost of estimating the kinetic term $T=\sum_{\alpha=1}^d T_{\alpha}$. Since each $T_\alpha$ is a dense matrix of size $L$ where $L$ is the number of grid points, we need $L^2$ independent circuits to compute $\langle \psi | T_\alpha |  \frac{\partial \psi}{\partial \theta_i} \rangle $. Fortunately the kinetic terms from different dimensions are decoupled, thus $T$ is sparse in higher dimensional systems with only $dL^2$ non-trivial terms, and the number of quantum circuits required for computing $\langle \psi | T  |  \frac{\partial \psi}{\partial \theta_i} \rangle $ therefore scales linearly with dimensionality. 
For the entire force vector, we need $dL^2N_\theta$ independent circuits. 

The estimation of the $\langle \psi | V |  \frac{\partial \psi}{\partial \theta_i} \rangle$ in general requires $O(L^d)$ independent circuits since $V$ is a diagonal matrix with $L^d$  entries, i.e. 
\begin{eqnarray}
V = \sum_{l_1, l_2,\cdots,l_N = 0, z} \beta_{l_1l_2...l_N} \sigma_1^{l_1} \sigma_2^{l_2}...\sigma_N^{l_N},
\end{eqnarray}
where $\sigma^0$ denotes identity matrix.
For potential energy in which $\mbox{e}^{-iV t}$ can be efficiently implemented, we can employ the decomposition
\begin{eqnarray}
\bra{\psi} V \ket{\frac{\partial \psi}{\partial \theta_i}} = \lim_{\epsilon \to  0}\frac{i}{2\epsilon} \bra{\psi} \mbox{e}^{-i\epsilon V} -  \mbox{e}^{i\epsilon V} \ket{\frac{\partial \psi}{\partial \theta_i}},
\end{eqnarray}
and use extrapolation to obtain the $\epsilon=0$ limit. 
However to the best of our knowledge, efficient implementation of the quantum circuits for $\mbox{e}^{-iV t}$ are only known for several special cases~\cite{Benenti2008, Macridin2018, Ollitrault2020}.
In other words, if the system cannot be efficiently simulated using the split product formula in Eq.~(\ref{eq:trotter}), it also cannot be efficiently simulated using the real-time VQA since it requires $O(2^N)$ (note that $2^N = L^d$) quantum circuits to evaluate $\langle \psi | V |   \frac{\partial \psi}{\partial \theta_i} \rangle$. 

Fortunately, for the imaginary-time VQA used in the subspace method, we only need the real part of $f_i$, which is related to the gradient of the energy by
\begin{eqnarray}
Re[\langle \psi | H |  \frac{\partial \psi}{\partial \theta_i} \rangle] = \frac{1}{2} \frac{\partial}{\partial \theta_i} \langle  \psi | H |  \psi \rangle. 
\end{eqnarray}
Since $V$ is a diagonal matrix and can be expressed in terms of $\sigma^z$ and identity matrices, we only need one circuit in the $z$-basis to obtain the expectation value $\langle \psi |V| \psi \rangle$. The evaluation of the momentum term $\langle \psi |T| \psi \rangle$ would again require $dL^2$ circuits. The gradient $\frac{\partial}{\partial \theta_i} \langle  \psi | H |  \psi \rangle$ can then be obtained either by finite difference method or the analytical parameter-shift rule~\cite{schuld2019evaluating, Mitarai2018}. Thus the combined measurement cost for the subspace method with imaginary time VQA scales as $O(N_\theta^2 + N_\theta d  L^2)$.

For the imaginary-time VQA, computing the matrix $M$ could still require a large number of measurements, in the order of $N_\theta^2 \propto N^4 =  (d\log_{2}L)^4$ if one uses HVA with up to two-qubit rotations. An alternative strategy is to simply use gradient descent in the optimization step when computing the ground states, though imaginary time algorithm has been shown to outperform gradient descent~\cite{McArdle2019}.

\section{Numerical Examples} \label{sec:numerics}
\subsection{One-Particle Problem: Isomerization Reaction Dynamics of Malonaldehydes}

We first demonstrate the capability of VQA in simulating chemical dynamics by considering a one-dimensional model of laser-driven isomerization reaction, namely, the hydrogen-transfer reaction of nonsymmetric substituted malonaldehydes~\cite{Doslic1998}. This chemical reaction has previously been simulated experimentally using the NMR quantum simulator~\cite{Lu2011}.

\begin{figure}[ht!]
    \centering
  \includegraphics[width=5.0in]{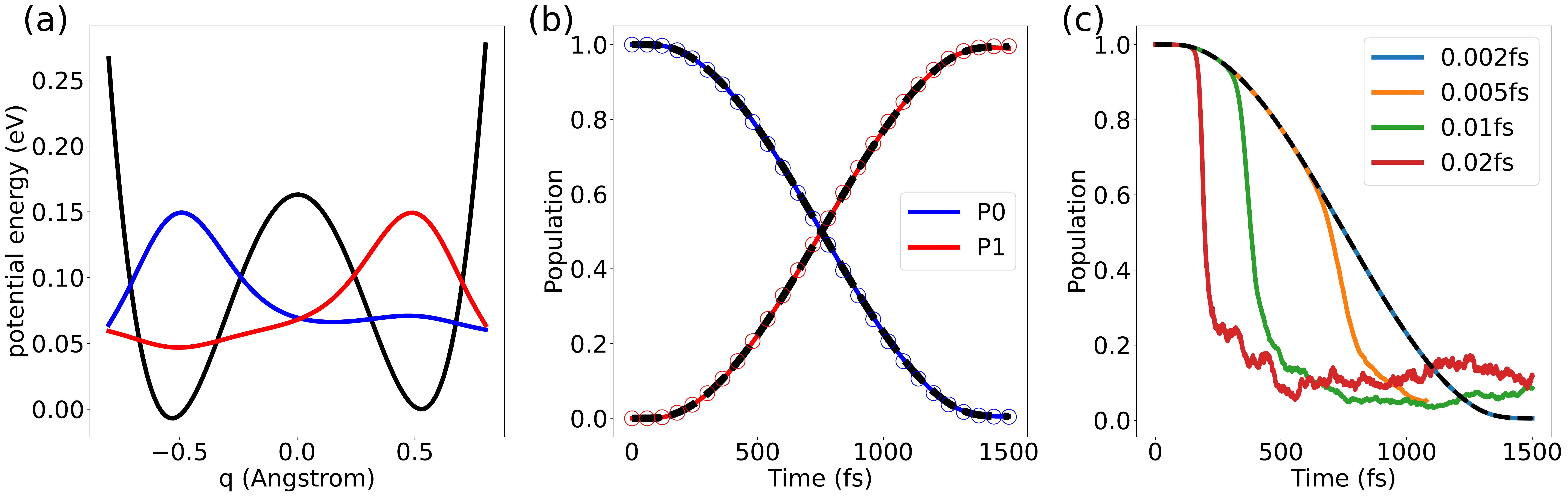}
    \caption{ Isomerization reaction dynamics of malonaldehydes. 
    (a) Potential energy curve (black) of the double-well potential (i.e. Eq.~(\ref{eq:double_well})). The blue and red lines depict the ground and first excited state wavefunctions, respectively.  
    (b) Simulation results of 1D isomerization dynamics using VQAs. The ground state population (reactant state) is denoted as $P_0$ and the first excited state population (product state) is denoted as $P_1$. The black dashed lines are numerically exact results, the solid lines are the results from time-dependent VQA, and the symbols are the results from the subspace method using two lowest energy eigenstates.
    (c) Ground state population dynamics from real time VQA using different time step sizes, the dashed black line denotes numerically exact result.}
    \label{fig:1D_VQA}
\end{figure}

The total Hamiltonian in the presence of an external driving field is given by 
\begin{eqnarray}
H = T + V  + H_I(t), \,\,\,\, H_I(t) = - \mu \varepsilon(t),
\end{eqnarray}
where $\mu = e x$ is the dipole moment operator, $x$ is the reaction coordinate and $\varepsilon(t)$ is the external driving field. The potential energy term is a tilted double-well potential
\begin{eqnarray} \label{eq:double_well}
V = \frac{\Delta}{2q_0}(x - x_0) + \frac{V^\ddagger-\Delta/2}{x_0^4}(x - x_0)^2(x + x_0)^2,
\end{eqnarray}
where $V^\ddagger$ is the barrier height, $\Delta$ denotes the asymmetry of the two wells, and $\pm x_0$ give the locations of the potential well minima. The values of parameters in atomic units are taken from Ref.~\citenum{Doslic1998}: $V^\ddagger=0.00625 $, $\Delta = 0.000257 $, $x_0 = 1$. 
The initial reactant state is assumed to be the ground state of the unperturbed Hamiltonian, which is mainly localized on the left well (blue line in Fig.~\ref{fig:1D_VQA}). We denote the reactant state population to be $P_0$. The first excited state is mainly localized in the right potential well, and we consider it to be the product state and denote its population as $P_1$.

The shape of the laser pulse is given by~\cite{Doslic1998}
\begin{equation}
\label{eq:laser_shape}
 \varepsilon(t) = 
    \begin{cases} 
      \varepsilon_0 \sin^2(\pi t / 2 s_1), & 0\le t\le s_1 \\
      \varepsilon_0, & s_1 <  t < s_2 \\
      \varepsilon_0 \sin^2(\pi (t_f-t) / 2 (t_f-s_2)), & s_2\le t\le t_f 
    \end{cases}
\end{equation}
where $t_f$ = 1500 fs, $s_1$ = 150 fs, $s_2$ = 1250 fs, and $\varepsilon_0$ = 0.00137 a.u. 

In our simulations, the system is discretized into 8 grid points between $\pm 0.8$ \AA~ such that the wavefunction can be encoded into $3$ qubits. The numerical results using the real time VQA and the subspace expansion method are presented in Fig.~\ref{fig:1D_VQA} (b), and it can be seen that both approaches provide good agreement with numerically exact results (black dashed lines). However, it is worth noting that the real-time algorithm requires a very small time step of $\delta t = 0.002$fs in order to attain good results.  
To illustrate the dependence of the performance of the real-time VQA on step size, we perform simulations of the isomerization dynamics using different step sizes, and the results are shown in Fig.~\ref{fig:1D_VQA} (c). It can be seen that the algorithm can become unstable quickly unless a very small time step is used. We found that the step size needs to be as small as $\delta t = 0.002$ fs to achieve good agreement with exact results, which translates to $750000$ iterations in total for the entire simulation in Fig.~\ref{fig:1D_VQA}.
In comparison, in the subspace method, we only use $1000$ iterations in the imaginary-time algorithm to find each eigenstate. To obtain the results in Fig.~\ref{fig:1D_VQA}(b) we only include two eigenstates in the subspace method since the other eigenstates are energetically well separated from these two states. 

\subsection{Two-Particle Problem: One-Dimensional Helium Atom}
Here we demonstrate our method with a more complicated example involving two particles, a one-dimensional model of helium atom. The electronic Hamiltonian with soft Coulomb potentials in atomic unit is given by
\begin{eqnarray}
H_0 =  \frac{1}{2} p_x^2 + \frac{1}{2} p_y^2 -  \frac{2}{\sqrt{x^2 + a^2}} -  \frac{2}{\sqrt{y^2 + a^2}} + \frac{1}{\sqrt{(x-y)^2 + a^2}},
\label{eq:he_ham}
\end{eqnarray}
where $x$, $y$ and $p_x$, $p_y$ are the positions and momenta of each electron, respectively, and $a$ is the softening parameter. Despite its simplicity, the one-dimensional helium model has been used in many previous works, e.g., to identify double ionization mechanism~\cite{Lein2000, Liao2012}, to probe Fano resonances~\cite{Zhao2012}, to investigate attosecond spectroscopy~\cite{Gaarde2011,Yang2015}, to study high-order harmonic generation~\cite{Erhand1997,Shi2021}, and to propose a new molecular imaging method~\cite{VanDerZwan2012}, just to name a few. In the following, we will demonstrate the capability of the subspace expansion approach in computing the harmonic generation spectrum for this model.

The model helium atom is initially in the ground state and subject to an intensive infrared laser pulse with a trapezoidal profile, given by~\cite{Shi2021}
\begin{equation}
\varepsilon(t) = 
    \begin{cases} 
      \varepsilon_0 \sin^2 \left ( \frac{\pi t}{4 T} \right )  \cos(\omega t), & 0 \le t \le 2T \\
      \varepsilon_0  \cos(\omega t), & 2T \le t \le 10T \\
      \varepsilon_0 \cos^2 \left ( \frac{\pi t}{4 T} - \frac{5 \pi }{2 T} \right ) \cos(\omega t), & 10T < t \le 12T 
    \end{cases}
\end{equation}
where $\varepsilon_0$ and $\omega$ are the electric field amplitude and frequency, respectively, and $T = 2\pi / \omega$ is the period of the electric field. Following Ref. \citenum{Shi2021}, the laser intensity is chosen to be $I =| \varepsilon_0| ^2 = 3 \times 10^{12}\; W/cm^2$, and the laser frequency is $\omega =0.3542$ eV (namely, $T = 11.67$ fs, and the wavelength is 3500 nm). We discretize the system into 8 grid points between $\pm 2.0$ \AA~ in each dimension, thus the wavefunction can be encoded into 6 qubits. The softening parameter in Eq. (\ref{eq:he_ham}) is chosen to be 0.7397$a_0$ so that the ground-state energy of the model roughly matches the experimental binding energy of helium.\cite{Kramida2020} 
Within the dipole approximation, the light-matter interaction Hamiltonian in the length gauge is given by 
\begin{equation}
    H_I(t) = \varepsilon(t)(x+y).
\end{equation}
The harmonic generation spectrum can be computed in the dipole form as the squared modulus of the Fourier transform of the time-dependent dipole moment over the pulse duration,\cite{Schultz2014,Erhand1997,Tong1998,Wozniak2021}
 \begin{eqnarray} \label{eq:hhg_fourier}
    I(\omega) \sim \left |\int^{12T}_0 d(t) \mbox{e}^{i\omega t} dt \right |^2 \label{eq:helium_fourier},
 \end{eqnarray}
where the time-dependent dipole is given by
\begin{eqnarray}
    d(t) = - \bra{\psi(t)} x + y \ket{\psi(t)},
    \label{eq:helium_dipole}
\end{eqnarray}
and $\ket{\psi(t)}$ is the electronic wavefunction of the model helium at time $t$. We numerically integrated the Schr\"{o}dinger equation in the DVR basis using a time step of 0.58 fs, and the obtained results are considered to be ``exact". For the subspace expansion method, 6 lowest eigenstates out of a total of 64 states were included in the low-energy subspace, and the same time step was used for the time evolution of the system. We do not observe noticeable difference when a smaller time step is used. 
 
\begin{figure}[ht!]
    \centering
  \includegraphics[width=0.9\textwidth]{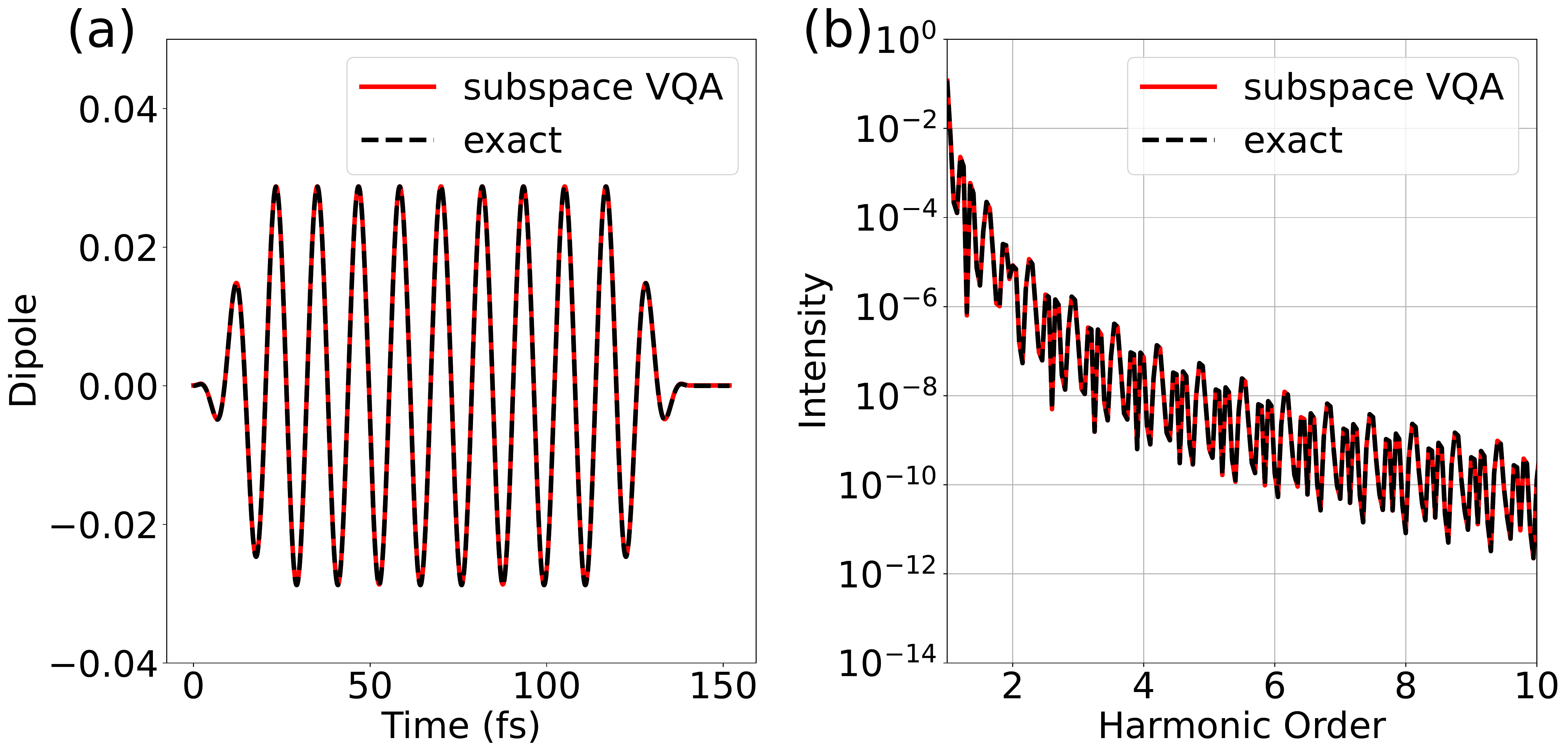}
    \caption{Helium atom: the dashed black lines are numerically exact results whereas solid red lines are results obtained from the subspace expansion method using 6 lowest eigenstates. 
    (a) The time dependence of dipole moment (see Eq. \ref{eq:helium_dipole}) under the driving field.
    (b) Harmonic generation spectrum obtained using Eq.~(\ref{eq:hhg_fourier}).}
    \label{fig:2D_subspace}
\end{figure}

The time evolution of the dipole moment and its corresponding harmonic generation spectrum computed using the low-energy subspace method are shown in Fig.~\ref{fig:2D_subspace}.
It can be seen that the results from the subspace method is in excellent agreement with the numerically exact simulations, further confirming the capability of our approach. This agreement is in fact extended up to all the harmonics within the frequency range limited by our time step (approximately 30th harmonics).

\newpage
\section{Discussions and Conclusions} \label{sec:discussion}
In this work we present a framework based on the DVR basis to perform variational quantum simulation of chemical dynamics. The use of DVR basis in representing a real-space Hamiltonian offers several advantages: (1) it requires no numerical overlap integral evaluation since simple analytical expression is available, (2) the resulting Hamiltonian is sparse and (3) fast convergence with respect to the number of grid points~\cite{Colbert1992}.
With the binary encoding, a $d$-dimensional system with $L$ grid points per dimension can be mapped into $d \log_2(L)$ qubits. For example, a 12-dimensional system with 32 grid points in each dimension can be encoded onto 60 qubits, a qubit number already present in the superconducting platform~\cite{Wu2021, ibm_roadmap}, though the error rates are still too big to perform non-trivial quantum simulations. On the other hand, storing and operating vectors and matrices with $32^{12}\approx 1.2\times10^{18}$ dimensions is a formidable task for classical computers. Such system sizes are challenging even for the  standard MCTDH unless additional approximations are made~\cite{Meyer2011, Manthe2017}. 
While we only consider systems in Cartesian coordinates in this work, the DVR framework is also applicable in polar and radial coordinates, useful for systems with rotational degree of freedom.

Measurement cost is a major concern for many VQAs, particularly for quantum dynamics simulations~\cite{Miessen2021, Lee2020a, Lee2021, Lee2021a}. Indeed we show that direct application of the real-time VQA would require a large number of circuit evaluations for two reasons: (1) a very small time step is needed in solving the differential equation in Eq.~(\ref{eq:EOM_theta}) and this problem deteriorates with system sizes, and (2) computing Im$[f]$ in Eq.~(\ref{eq:EOM_theta}) is in general not efficient for arbitrary potential energy, i.e. the number of quantum circuits grows exponentially with qubit number. These drawbacks limit the application of the real-time VQA for simulations of chemical dynamics. Motivated by the intuition that many chemical dynamics involve only low-energy states, we propose expressing system Hamiltonian in the low-energy subspace, a task that can be achieved by quantum computer with $O(N_\theta^2 + N_\theta d L^2)$ independent circuits if we use the imaginary-time VQA to find the eigenstates, recalling that $N_\theta$ is the number of variational parameters, $d$ is the dimensionality and $L$ is the number of grid points per dimension. 
The measurement cost reduces to simply $O(N_\theta d L^2)$ if we use gradient descent for optimizing the variational parameters. Despite the polynomial scaling with dimensionality, the number of circuits needed could still be a large number, one should therefore combine with optimal measurement strategies to further reduce the number of circuit evaluations~\cite{Wang2019, Hamamura2020, Gokhale2020, Crawford2021}.

The quantum advantage of the subspace method arises from two parts: (1) computing the ground state of the modified Hamiltonian in Eq.~(\ref{eq:excited_hamiltonian}), and (2) projecting the total Hamiltonian on the system eigenstate subspace in Eq.~(\ref{eq:projection}). The classical computational cost of both calculations scale exponentially with system dimensionality whereas the computational cost is only polynomial in quantum computers.
Additionally, a large body of works have been developed to find the ground or low-energy excited states of a Hamiltonian~\cite{Cerezo2021}, these methods can be easily adopted into our subspace approach. 
Furthermore the quality of the ground or excited state computation can generally be systematically improved by increasing the circuit depth in the ansatz or via the use of extra ancillary qubits~\cite{Hsieh2021}. 
 
Finding an effective low-energy Hamiltonian is an active field of research in theoretical physics with vast applications in many-body systems and beyond~\cite{sachdev1999quantum, Bravyi2011}. Here we adopt a relatively straightforward approach by projecting the Hamiltonian into the low-energy eigenstate sub-space. While many interesting physics takes place in the low-energy regime, there could be exceptions. For example, in complex systems with dense spectrum or processes involving multi-electron ionization, the number of eigenstates to be included in the subspace might be large which makes direct application of the subspace method difficult. For these systems, it might be possible to find a renormalized effective Hamiltonian that incorporates the influence of high-energy states using methods such as the Schrieffer-Wolff transformation~\cite{Bravyi2011}. A recent work shows that the use of effective low-energy Hamiltonians can result in lower errors in Hamiltonian simulation with Suzuki-Trotter decomposition~\cite{Sahinoglu2021}. 

To conclude, we present variational simulation of quantum dynamics for quantum systems in real space. Using the DVR basis and a binary encoding scheme, a $d$-dimensional system with $L$ grid points per dimension can be encoded onto $d\log_2(L)$ qubits. For time propagation, we show that direct application of real-time VQA based on the McLaclan’s principle is not efficient for general potential energy due to the large number of quantum circuits required. Motivated by the insight that many chemical dynamics only involve a small number of low-energy states, we propose projecting the system Hamiltonian into the low-energy eigenstate subspace with imaginary-time VQAs and classically computing the exact dynamics within the subspace. We show that such approach is efficient for general potential energy, with the number of quantum circuits scaling only polynomially with dimensionality. Our numerical examples show that the low-energy subspace expansion approach is capable of representing the true quantum dynamics with high accuracy, even in the presence of intense driving fields. 
Our work opens up the possibility of simulating the dynamics of quantum systems in real space, e.g. atomic and molecular systems, with near-term quantum computers. 

\section{acknowledgement}
L.S. acknowledges the support from the University of California Merced start-up funding. 

\newpage
\appendix

\section{Computing $M$ and $f$}
\begin{figure}[ht!]
    \centering
  \includegraphics[width=0.9\textwidth]{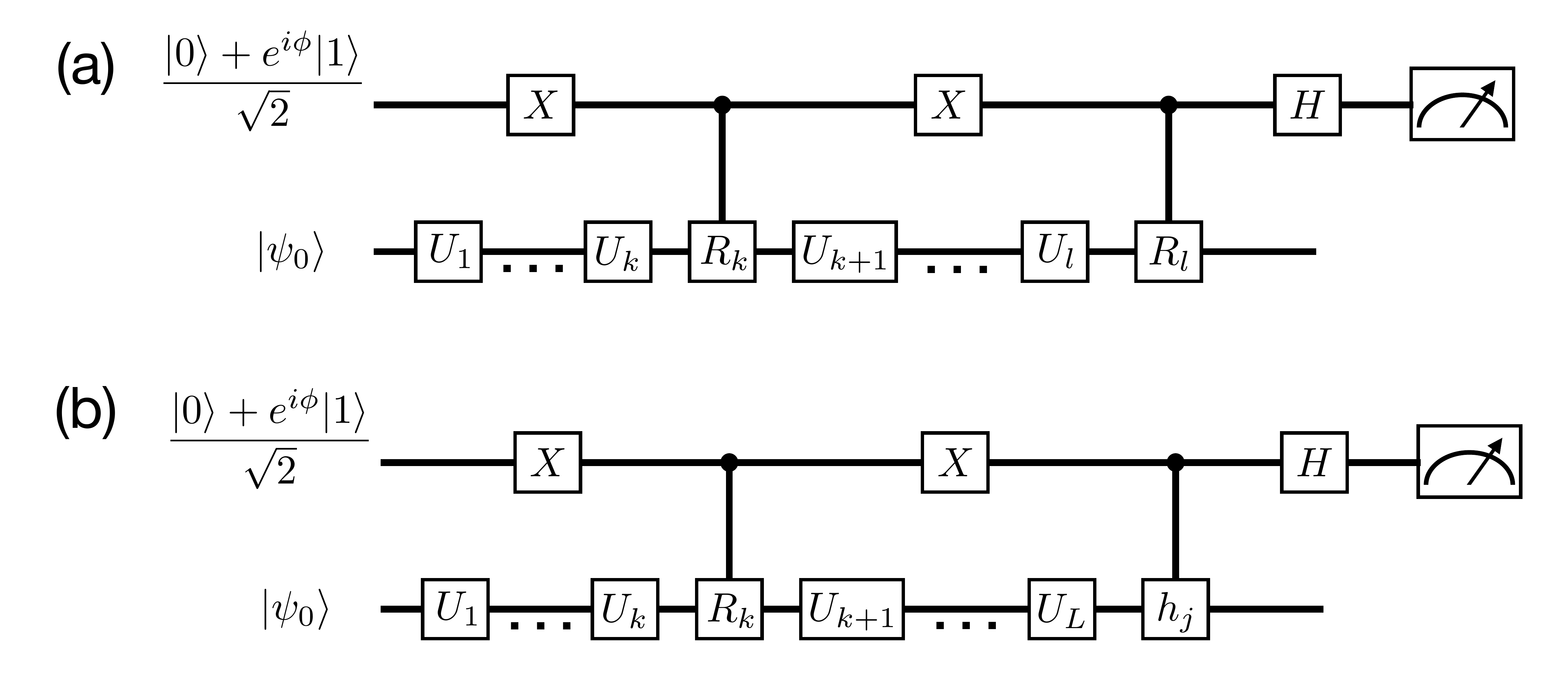}
    \caption{Quantum circuits to compute the matrix elements (a) $M_{kl}$ and (b) $ f_k$ in Eq.~(\ref{eqn:circuit}). The ancillary qubit is initialized in state $\frac{\ket{0} + e^{i\phi} \ket{1} }{\sqrt{2}}$. The phase factor $\phi$ is set to be $0$ or $\pi/2$ in order to measure the real and imaginary components of the expectation values, respectively.}
    \label{fig:circuit}
\end{figure}

Here we show that the matrix elements in $ M$ and $ f$ of Eqs.~(\ref{eq:EOM_theta}) and (\ref{eq:EOM_theta_imag}) can be measured in a quantum computer. 
We note that the derivative of each term in Eq.~(\ref{eqn:wfn_ansatz}) is written as $\frac{\partial {U}_k(\theta_k)}{\partial \theta_k} = i {R}_k \hat{U}_k(\theta_k)$, therefore the matrix elements of $ M$ and $ f$ are (assuming $k<l$)
\begin{eqnarray} \label{eqn:circuit}
 M_{kl} &=&     \bra{\psi_0}  U_1^\dagger  ... U_k^\dagger  R_k^\dagger ... \hat U_L^\dagger
                         U_L^ ...  R_l  U_l ...  U_1  \ket{\psi_0}  \\
       &=&  \big( \bra{\psi_0}  U_1^\dagger  ... U_k^\dagger  R_k^\dagger  U_{k+1}^\dagger  ... U_l^\dagger  R_l  U_l ...  U_1  \ket{\psi_0} \big) ,  \nonumber \\
 f_{k}   &=&   i \sum_j c_j \bra{\psi_0}  U_1^\dagger  ... U_L^\dagger  h_j
                         U_L^ ...  R_k  U_k ...  U_1  \ket{\psi_0} ,   \nonumber         
\end{eqnarray}
where we have expressed the Hamiltonian as $ H = \sum_j c_j {h}_j$ where $h_j$ are some Pauli strings and $c_j$ are the corresponding coefficients. 
The real and imaginary parts of the matrix elements in Eq. (\ref{eqn:circuit}) can be obtained via the Hadamard test, and the structures of the circuits are shown in Fig.~\ref{fig:circuit}.

\clearpage
\bibliographystyle{naturemag_noURL.bst}
\bibliography{MyCollection}

\end{document}